\documentclass[a4paper,preprintnumbers,floatfix,superscriptaddress,pra,twocolumn,showpacs,notitlepage,longbibliography, aps]{revtex4-2}

\usepackage[utf8]{inputenc}
\usepackage[T1]{fontenc}
\usepackage{amsmath, amsthm, amssymb,amsfonts,mathbbol,amstext}
\usepackage{graphicx}
\usepackage{dcolumn}
\usepackage{bm}
\usepackage{bbm}
\usepackage{hyperref}
\usepackage{mathtools}
\usepackage{comment}
\usepackage{color}
\usepackage{multirow}
\usepackage{pgf,tikz}
\usepackage{mathrsfs}
\usepackage{physics}
\usetikzlibrary{arrows}
\usepackage{soul,xcolor}
\usepackage{adjustbox}
\usepackage{dsfont}
\usepackage{amssymb}

\begin{document}





\title{The Apparatus Strikes Back: Momentum Conservation and the  Cost of Spatial Superpositions}

\author{Lucas C. C\'eleri\href{https://orcid.org/0000-0001-5120-8176}{\includegraphics[scale=0.05]{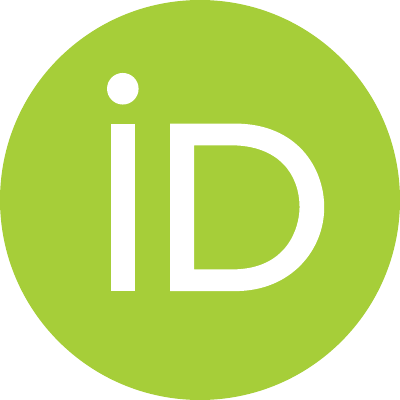}}}
\email{lucas@qpequi.com}
\affiliation{QPequi Group, Institute of Physics, Federal University of Goi\'as, Goi\^ania, Goi\'as, 74.690-900, Brazil}
\affiliation{S\~ao Carlos Institute of Physics, University of S\~ao Paulo, IFSC-USP, 13566-590, S\~ao Carlos, SP, Brazil}

\author{Diogo O.\ Soares-Pinto\href{https://orcid.org/0000-0002-4293-6144}{\includegraphics[scale=0.05]{orcidid.pdf}}}
\email{dosp@ifsc.usp.br}
\affiliation{S\~ao Carlos Institute of Physics, University of S\~ao Paulo, IFSC-USP, 13566-590, S\~ao Carlos, SP, Brazil}

\author{Daniel A.~Turolla Vanzella\href{https://orcid.org/0000-0003-2512-1290}{\includegraphics[scale=0.05]{orcidid.pdf}}}
\email{vanzella@ifsc.usp.br}
\affiliation{S\~ao Carlos Institute of Physics, University of S\~ao Paulo, IFSC-USP, 13566-590, S\~ao Carlos, SP, Brazil}

\begin{abstract}
Preparing massive particles in coherent spatial superpositions is a central objective of modern quantum science, motivated by applications ranging from fundamental tests of quantum mechanics and gravity to quantum-enhanced sensing. The experimental difficulty of realizing such superpositions is usually attributed to environmental decoherence mechanisms whose impact depends on the specific details of the experimental implementation. Here we identify a universal constraint arising from momentum conservation and the quantum nature of the preparation apparatus. We show that any protocol that places a particle of mass $m$ in a spatial superposition 
with separation $d$ necessarily entangles the particle with the center-of-mass degree of freedom of the apparatus responsible for the splitting. The resulting recoil displacement, fixed by conservation of the mass dipole moment, reduces the coherence of the particle when the apparatus is not explicitly included as part of the quantum system. For an apparatus of mass $M$, the preservation of coherence requires the recoil displacement to remain smaller than the coherence length of the apparatus center-of-mass state, 
leading to a quantitative bound that can be expressed as a constraint on 
its temperature and  on how rigidly the apparatus is anchored to the laboratory frame. 
We analyze the implications of this bound for current matter-wave interferometry experiments, proposals for gravitationally mediated entanglement, and tests of quantum mechanics near the Planck scale.
Our main result, which may seem counterintuitive, is that the recoil of even heavy macroscopic apparatuses can pose a surprisingly strong constraint on 
the coherence of spatial superpositions of particles with masses
well below the Planck mass.
Finally, we discuss why the resulting limitation should be regarded as a conservative estimate and under which conditions it can be interpreted as an instance of false decoherence.
\end{abstract}

\maketitle


The ability to prepare increasingly massive systems in coherent spatial superpositions is motivated by a wide range of scientific questions that span several disciplines. From a fundamental perspective, large-mass interferometry provides one of the most direct probes of the validity of the quantum superposition principle in regimes far beyond those tested by conventional atomic and molecular experiments, allowing experimental tests of collapse models and other proposed modifications of quantum mechanics~\cite{Bassi2013,Bassi2023}. Closely related is the possibility of probing the interface between gravity and quantum theory, where massive superpositions may reveal genuinely quantum features of the gravitational interaction, as in the Bose–Marletto–Vedral (BMV) proposal~\cite{Bose2017,Marletto2017} (see also Ref.~\cite{Carney2019}). Beyond foundational questions, coherent superpositions constitute valuable resources for quantum-enhanced sensing and metrology, promising unprecedented sensitivities in measurements of forces, accelerations, rotations, and gravitational fields~\cite{Degen2017}. Increasing the complexity and mass of quantum objects also raises intriguing questions concerning the quantum-to-classical transition and the ultimate limits of quantum coherence in complex systems, including biological matter and, in the long term, possibly even living organisms~\cite{Isart2010}. Together, these motivations have made the creation of large-scale quantum superpositions an important objective of contemporary quantum science.

Realizing such proposals remains experimentally demanding. At the mass scales currently targeted by matter-wave interferometry experiments, the dominant theoretical and experimental limitations are usually associated with environmental decoherence channels such as scattering by thermal photons~\cite{RomeroIsart2011}, gas collisions, phonon excitation~\cite{Henkel2022,Folman2024}, gravitational radiation~\cite{Satishchandran2024}, gravitational decoherence induced by the apparatus itself~\cite{Gunnink2023}, and acceleration noise~\cite{Grossardt2020}. Each of these mechanisms is based on a specific physical process and gives rise to constraints that depend on the geometry, materials, and dynamical details of the experimental implementation.

In this article, we show that, independently of these environmental mechanisms, any preparation scheme for a massive spatial superposition must satisfy an additional constraint originating solely from momentum conservation and the quantum nature of the preparation apparatus. The argument makes an elementary physical observation precise: whenever a particle is
displaced, the apparatus responsible for generating the 
displacement must recoil. In a quantum-mechanical description, this recoil is itself subject to uncertainty and therefore generically entangles the particle with the apparatus. The resulting loss of coherence in the reduced state of the particle is not associated with any particular environmental channel, but follows directly from local conservation of momentum together with the canonical commutation relation between position and momentum.

This mechanism is closely related to the Einstein-Bohr recoiling-slit \emph{Gedankenexperiment}, as reconstructed by Bohr~\cite{BohrEinstein} and later given quantitative form by Wootters and Zurek~\cite{WoottersZurek1979}. In that classic analysis, a slit that recoils when traversed by a photon acquires a path-dependent quantum state, and interference visibility is reduced whenever the recoil renders the two slit states distinguishable. Recent experiments have realized this scenario in a controllable form using a single trapped atom as a recoiling slit and have verified complementarity at the single-quantum level~\cite{Liu2024recoiling}. Closely related physics also appears in the theory of environment-induced decoherence~\cite{JoosZeh1985}, where information about a quantum system is encoded in auxiliary degrees of freedom and becomes inaccessible when these are ignored.

The situation considered here differs from the recoiling-slit scenario in two important respects. First, the recoiling object is not a microscopic quantum system but a macroscopic apparatus whose center-of-mass degree of freedom is generally in a thermal state. Second, the relevant conservation law is not the momentum transfer associated with a specific interaction but the conservation of the mass dipole moment of the combined particle-apparatus system. As a consequence, the resulting constraint is independent of the mechanism used to generate the superposition and applies equally to optical, magnetic, electromagnetic, mechanical, or any other splitting protocol. By combining mass-dipole conservation with the canonical commutation relation between position and momentum, we derive a quantitative bound on the coherence that can be retained in the reduced state of the particle. The bound depends only on a small set of macroscopic parameters, the particle mass, the 
separation
of the superposed position states, and the mass and center-of-mass state of the apparatus, which
can be expressed as a temperature constraint on the apparatus center-of-mass mode. In this sense, we identify a {\it universal} lower bound on the experimental cost of preparing massive quantum superpositions. 

\section{Results}
\noindent
\textbf{Momentum conservation and coherence} 

\noindent Consider a particle of mass $m$ initially prepared in a state $\ket{\psi_0}$ localized around the point $\vec{x}_0 = \expval{\hat{\vec{x}}}{\psi_0}$, with position uncertainty characterized by $\sigma_0^2 = \expval{\hat{\vec{x}}^{\,2}}{\psi_0}-\vec{x}_0^{\,2}$. An auxiliary system, hereafter referred to as the apparatus, is initially at rest and interacts with the particle to prepare it in a superposition of two spatially separated states. These states are taken to be translations of $\ket{\psi_0}$ by $\mp \vec d/2$,
\begin{equation}
\ket{\psi_L} = \hat{U}_{L}^{(S)} \ket{\psi_0},
\qquad
\ket{\psi_R} = \hat{U}_{R}^{(S)}\ket{\psi_0},
\label{eq:pLR}
\end{equation}
where $\hat{U}_{L}^{(S)} = \hat{U}_{R}^{(S) \dagger} = e^{i\hat{\vec p}\cdot \vec d/(2\hbar)}$, $\hat{\vec p}$ is the particle momentum operator and $d =\|\vec d\|\gg \sigma_0$, so that $\ip{\psi_L}{\psi_R}\approx 0$. Ideally, one would like to prepare the coherent superposition $\ket{\psi} = \left(\ket{\psi_L} + \ket{\psi_R}\right)/\sqrt{2}$. As we will show,  momentum conservation imposes a constraint on this preparation procedure.

To see this, recall that the mass dipole moment of a collection of masses $m_j$, located at positions $\vec{x}_j$, is defined as $\vec{\mu} = \sum_j m_j\,\vec{x}_j = \mathcal{M}_{\rm tot}\,\vec{X}_{\rm cm}$, where $\mathcal{M}_{\rm tot} = \sum_j m_j$ is the total mass and $\vec{X}_{\rm cm}$ denotes the center-of-mass (CM) position. For an isolated system, momentum conservation implies conservation of the mass dipole moment in the CM rest frame, $\dot{\vec{\mu}} = \mathcal{M}_{\rm tot}\,\dot{\vec{X}}_{\rm cm} = \vec{P}_{\rm tot}\big|_{\rm CM} \equiv \vec{0}$, where the overdot denotes a time derivative and $\vec{P}_{\rm tot}\big|_{\rm CM}$ is the total momentum measured in the CM frame.

Suppose now that the joint system consists solely of a particle and an apparatus of total mass $M$. If the preparation procedure displaces the particle by $\pm\vec d/2$, conservation of the mass dipole moment requires a compensating displacement $\pm\vec D/2$ of the apparatus CM such that
\begin{equation}
m\,\vec d + M\,\vec D = \vec 0.
\label{eq:mass-dipole-balance}
\end{equation}
Consequently, even in the most favorable scenario in which the preparation affects only the CM degree of freedom of the apparatus, leaving its internal state unchanged, the particle necessarily becomes entangled with the apparatus. 

Consider the initial state of the system-apparatus as $\hat{\rho}_{in} = \dyad{\psi_0}{\psi_{0}} \otimes \hat{\rho}_0$, where $\ket{\psi_0}$ given in Eq.(\ref{eq:pLR}) is the initial state of the particle and $\hat{\rho}_0$ denotes the initial CM state of the apparatus. The resulting joint state after the system-apparatus interaction is given by
\begin{eqnarray}
\hat{\rho} &=& \frac{1}{2}\sum_{k,k'\in\{L,R\}} U_{k}^{(S)} \otimes U_{\bar k}^{(A)} \hat{\rho}_{in} \,U_{k^{\prime}}^{(S) \dagger} \otimes U_{\bar{k^{\prime}}}^{(A) \dagger}\nonumber \\
&=& \frac{1}{2} \sum_{k,k'\in\{L,R\}} \dyad{\psi_k}{\psi_{k'}} \otimes \hat{U}_{\bar{k}}^{(A)} \,\hat{\rho}_0\, \hat{U}_{\bar{k'}}^{(A) \dagger},
\label{eq:Ent}
\end{eqnarray}
where $\bar{k}$ is the ``complement'' of $k$ in $\{L,R\}$ --- i.e., $\bar{L} :=R$ and $\bar{R}:=L$ --- and
\begin{equation}
\hat{U}^{(A)}_L =\hat{U}^{(A) \dagger}_R = e^{-i\hat{\vec P}\cdot\vec D/(2\hbar)} = e^{im\hat{\vec P}\cdot\vec d/(2\hbar M)},
\label{eq:U}
\end{equation}
with $\hat{\vec P}$ being the momentum operator of the apparatus. Note that
the branches associated with $\ket{\psi_L}$ and $\ket{\psi_R}$ are correlated with opposite displacements of the apparatus CM.

Taking the partial trace over the degree of freedom of the apparatus yields the reduced state of the particle,
\begin{equation}
\hat{\rho}_S = \frac{1}{2} \sum_{k,k'\in\{L,R\}} \dyad{\psi_k}{\psi_{k'}}\,{\rm tr_{A}}\!\left(\hat{U}_{\bar{k}}^{(A)}\,\hat{\rho}_0\,\hat{U}_{\bar{k'}}^{(A) \dagger}\right).
\label{eq:rhop}
\end{equation}
Now, we have ${\rm tr_{A}}\![\hat{U}^{(A)}_{k}\,\hat{\rho}_0\,\hat{U}^{(A) \dagger}_{k}] = 1$, while for $k\ne k'$, and using the fact that $\hat{U}^{(A)}_R = \hat{U}_L^{(A) \dagger}$, we define the coherence factor
\begin{equation}
\chi_{\rm c} = {\rm tr} \!\left(\hat{\rho}_0 \hat{U}^{(A) 2}_R\right) = \left[{\rm tr}\!\left(\hat{\rho}_0\hat{U}_L^{(A) 2}\right)\right]^*.
\end{equation}
Equation~\eqref{eq:rhop} shows that the coherence of the reduced particle state is entirely controlled by $\chi_{\rm c}$, which quantifies the distinguishability of the recoil states of the apparatus. Indeed, the relative entropy of coherence~\cite{Baumgratz2014}, which measures the quantum coherence of the system, is given by
\begin{equation}
    C_{{\rm rel}} = \ln 2 - h_2\!\left(\frac{1+\abs{\chi_{\rm c}}}{2}\right),
\end{equation}
with $h_2(p) = -p\ln p - (1-p)\ln(1-p)$ being the binary entropy, 
which, in the present case, quantifies the correlations established between the particle and the apparatus. Determining this quantity is therefore essential for assessing the coherence retained by the reduced particle state.

In order to explicitly evaluate $\chi_{\rm c}$, some minimal modeling of the apparatus is required. For a realistic many-body system, it is convenient to separate the CM degree of freedom from the internal relative coordinates of its constituents. The previous analysis 
assumes that only the CM degree of freedom participates in the preparation of the states $\ket{\psi_L}$ and $\ket{\psi_R}$. As will be discussed below, this corresponds to the most favorable scenario. In general, excitation of internal degrees of freedom introduces additional sources of entanglement with the particle and therefore imposes more stringent requirements for the preparation of a coherent spatial superposition. The bound derived here should therefore be regarded as a necessary, but not sufficient, condition for a successful preparation protocol.

For concreteness, we assume that the CM of the apparatus is in thermal equilibrium at temperature $T$ and is confined by an effective harmonic potential with angular frequency $\omega$. This confinement accounts for the fact that, at some level, the apparatus is mechanically connected to the laboratory. As we shall see, a more rigidly anchored apparatus (larger $\omega$) makes the preparation of coherent superpositions of massive particles increasingly difficult. On the other hand, enlarging the portion of the experimental setup that participates coherently in the recoil increases the effective apparatus mass $M$, thus relaxing the corresponding constraints. The interplay between these two effects and the question of what should be regarded as ``the apparatus,'' will be addressed later.

Under these assumptions and defining the inverse thermal energy $\beta=1/(k_B T)$, the initial CM state of the apparatus is taken to be the Gibbs thermal state $\hat{\rho}_0 = e^{-\beta \hat H} / {\cal Z}$, with $\hat H = \hat{\vec P}^{\,2}/(2M) + M\omega^2\hat{\vec X}^{\,2}/2$ being the Hamiltonian, ${\cal Z} = {\rm tr}(e^{-\beta\hat H})$ is the partition function, and $\hat{\vec X}$ and $\hat{\vec P}$ denote the CM position and momentum operators of the apparatus, respectively. Using the position representation, the relation $\hat U_R^{(A)2}\ket{\vec X} = \ket{\vec X-\vec D}$, which follows from Eq.~\eqref{eq:U}, and the standard Mehler kernel~\cite{Feynman}, we obtain
\begin{equation}
\chi_{\rm c} = \exp\!\left[-\frac{D^2}{4 a_c}\coth\!\left(\frac{\epsilon}{2}\right)\right],
\label{eq:chi-derivation}
\end{equation}
where $\epsilon=\hbar\omega/(k_B T)$ is the energy ratio, $a_c = \hbar/(M\omega)$ is the effective area of the ground state of the CM degree of freedom of the apparatus, and we should recall 
that Eq.~(\ref{eq:mass-dipole-balance}) imposes the constraint $D=md/M$. At the quantum limit $\epsilon \gg 1$, this reduces to
\begin{equation}
\chi_{\rm c} \approx \exp\!\left[-\frac{D^2}{4 a_c}\right], 
\end{equation}
from which we clearly see that the factor $D/\sqrt{a_c}$ is a measure of the distinguishability of the displaced states of the apparatus. For the preparation of a coherent superposition of the particle, we need $D/\sqrt{a_c} =md\sqrt{\omega/(\hbar M)}\ll 1$, thus avoiding the entanglement between the particle and the apparatus.

\begin{figure*}[t]
  \centering
  \includegraphics[width=1\linewidth]{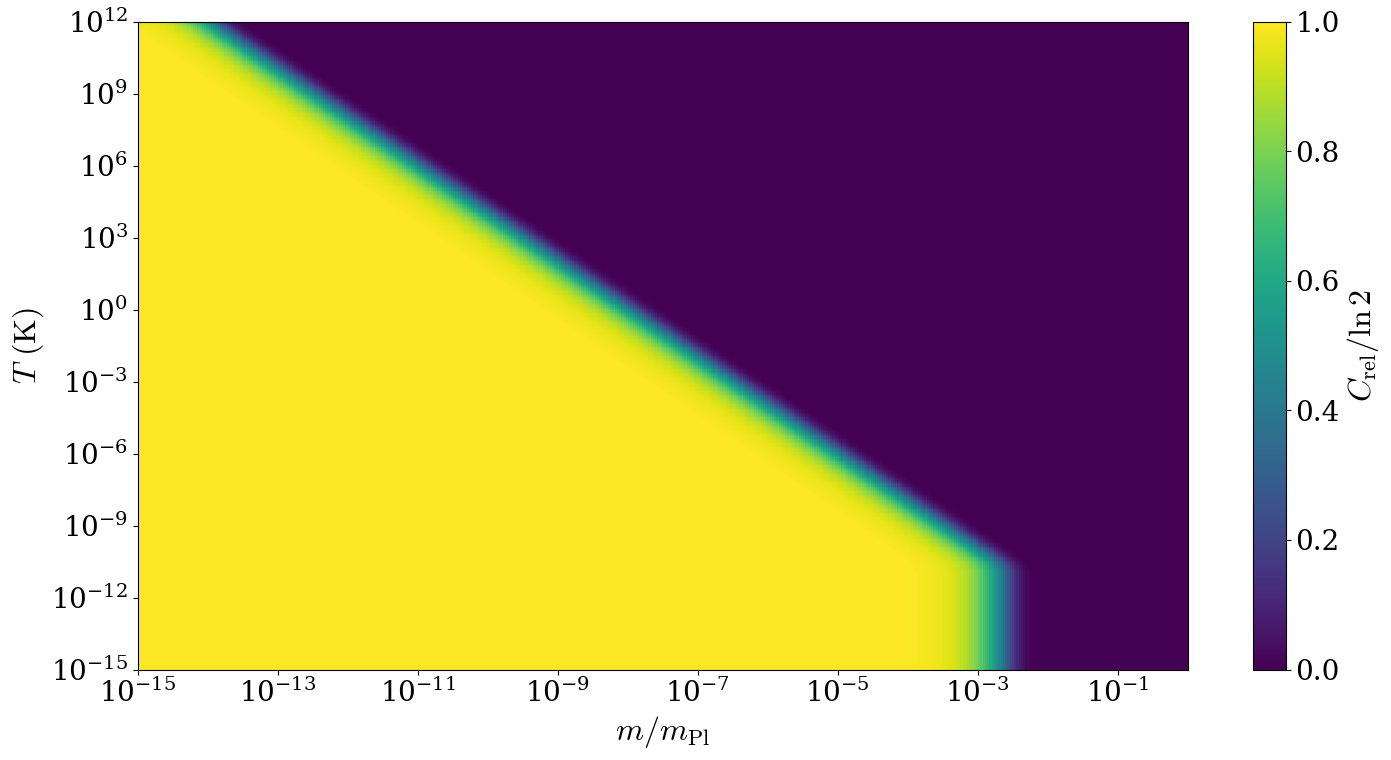}
 \caption{Normalized relative entropy of coherence, $C_{\rm rel}/\ln 2$, of the reduced state of the particle as a function of the particle mass, expressed in units of the Planck mass, and of the 
 temperature $T$. We fix $M=100$~kg, $d=1\,\mu{\rm m}$, and $\omega/(2\pi)=1$~Hz. The transition from the high-coherence (yellow) to the low-coherence (dark-blue) region reflects the increasing distinguishability of the apparatus recoil states associated with the two branches of the spatial superposition of the particle.  At any temperature, coherence decreases as the particle mass increases, 
 illustrating the growing impact of particle-apparatus entanglement imposed by momentum conservation. But notice also the existence
 of a critical mass value, $m_\ast \approx 3\times 10^{-3}~m_{\rm Pl}$, above which the coherence of the reduced particle state is strongly suppressed for all temperatures.}
  \label{fig:coherence_mT}
\end{figure*}

Figure~\ref{fig:coherence_mT} shows the normalized relative entropy of
coherence of the reduced state of the particle as a function of 
the temperature $T$
and of the particle mass expressed in units of Planck mass, $m_{\rm Pl}$. For any given temperature, the coherence decreases as the particle mass increases because momentum conservation requires a larger recoil of the apparatus, which in turn increases the entanglement between the particle and the apparatus CM degree of freedom (CM-DoF). Tracing out this DoF reduces the coherence of the particle according to Eq.~\eqref{eq:rhop}. Notice also that for the
parameters adopted in Fig.~\ref{fig:coherence_mT}, there
exists a critical mass value,  $m_\ast\approx 3\times 10^{-3}~m_{\rm Pl}$, above which the coherence of the reduced particle state 
is exponentially suppressed for {\it all} 
temperatures. As we shall see next, this is not an exception but
a generic feature.

\vspace{0.2cm}
\noindent \textbf{The bound on mass}

\noindent Equation~\eqref{eq:chi-derivation} shows that keeping the coherence of the reduced particle state requires the recoil displacement $D$ to remain smaller than the coherence length of the CM state. More precisely, condition $\chi_{\rm c}\sim 1$ implies that $D \lesssim \sigma_{\rm app}$, with $\sigma_{\rm app} = \sqrt{2a_c\tanh(\epsilon/2)}$. Combining this result with the momentum-conservation constraint, Eq.~\eqref{eq:mass-dipole-balance}, we arrive at the main result of this work:
\begin{equation}
md \lesssim M\sqrt{2a_c
\tanh\!\left(\frac{\epsilon}{2}\right)} = \sqrt{\frac{2\hbar M}{\omega}\tanh\!\left(\frac{\hbar\omega}{2k_B T}\right)}.
\label{eq:main-bound}
\end{equation}
This follows solely from momentum conservation, through the relation $D=md/M$, together with the canonical commutation relations encoded in the finite position uncertainty of the apparatus CM. Importantly, it is independent of the specific interaction employed to generate the spatial superposition.

We can identify from (\ref{eq:main-bound}) two distinct regimes:
\begin{eqnarray}
md \lesssim 
\left\{
\begin{array}{ll}
    \sqrt{2 \hbar M/\omega} &, \;\;  k_B T\ll \hbar \omega
    \\
    \\
    \sqrt{\hbar^2 M/(k_B T)} &,\;\;  k_B T\gg \hbar \omega
\end{array}
\right. .
\label{eq:main-bound-regimes}
\end{eqnarray}
For sufficiently low temperatures ($k_BT\ll \hbar \omega$), the apparatus CM approaches its ground state, for which the 
{\it coherence length} $\hbar/\sqrt{\langle \hat{P}^2\rangle}$ attains the maximum value allowed by the potential, $\sqrt{2a_c}$; note that the relevant scale here is the coherence length, not the position spread, which is in fact {\it smallest} in the ground state and grows with temperature.
The recoil 
states 
associated with the two branches 
 are then
nearly indistinguishable  provided the first line of
(\ref{eq:main-bound-regimes}) is respected --- which leads to the
temperature-independent condition 
\begin{equation}
    m\lesssim m_\ast :=\sqrt{2\hbar M/(\omega d^2)}.
\end{equation} 
This is represented by the yellow region  in the 
lower part of Fig.~\ref{fig:coherence_mT}, where
the reduced state of the particle retains a large fraction of its coherence.
In contrast, for large values of  $T$, thermal fluctuations increase the distinguishability of recoil states, leading to stronger particle-apparatus entanglement 
and
a rapid suppression of coherence, represented by the blue region
above the temperature-dependent
bound $m\lesssim \sqrt{\hbar^2 M/(k_B T d^2)}$ in 
Fig.~\ref{fig:coherence_mT}.

 A  neat way of rewriting the bound 
(\ref{eq:main-bound}), in terms of dimensionless ratios, is
\begin{eqnarray}
\frac{d}{\lambdabar_C} \lesssim\sqrt{\frac{Mc^2}{(\hbar \omega/2)}
\tanh\!\left(\frac{\hbar \omega}{2k_B T}\right)},
    \label{eq:main-bound-lambdaC}
\end{eqnarray}
where $\lambdabar_C :=\hbar/(mc)$ is the (reduced) Compton 
wavelength associated with the mass $m$ of the particle.
Notice that the r.h.s.\ of (\ref{eq:main-bound-lambdaC}) depends only
on the  apparatus, while the l.h.s.\ refers to the particle alone. Hence,
given the parameters of the apparatus, there is a bound on the separation
of the spatial
superposition of the particle, in units of its Compton wavelength, if coherence is
to be maintained without reference to the apparatus.
Any attempt to coherently 
superpose particle states
with separations larger than this value will fail
due to momentum conservation alone
if the apparatus is traced out.

\begin{figure*}[t]
  \centering
  \includegraphics[width=1\linewidth]{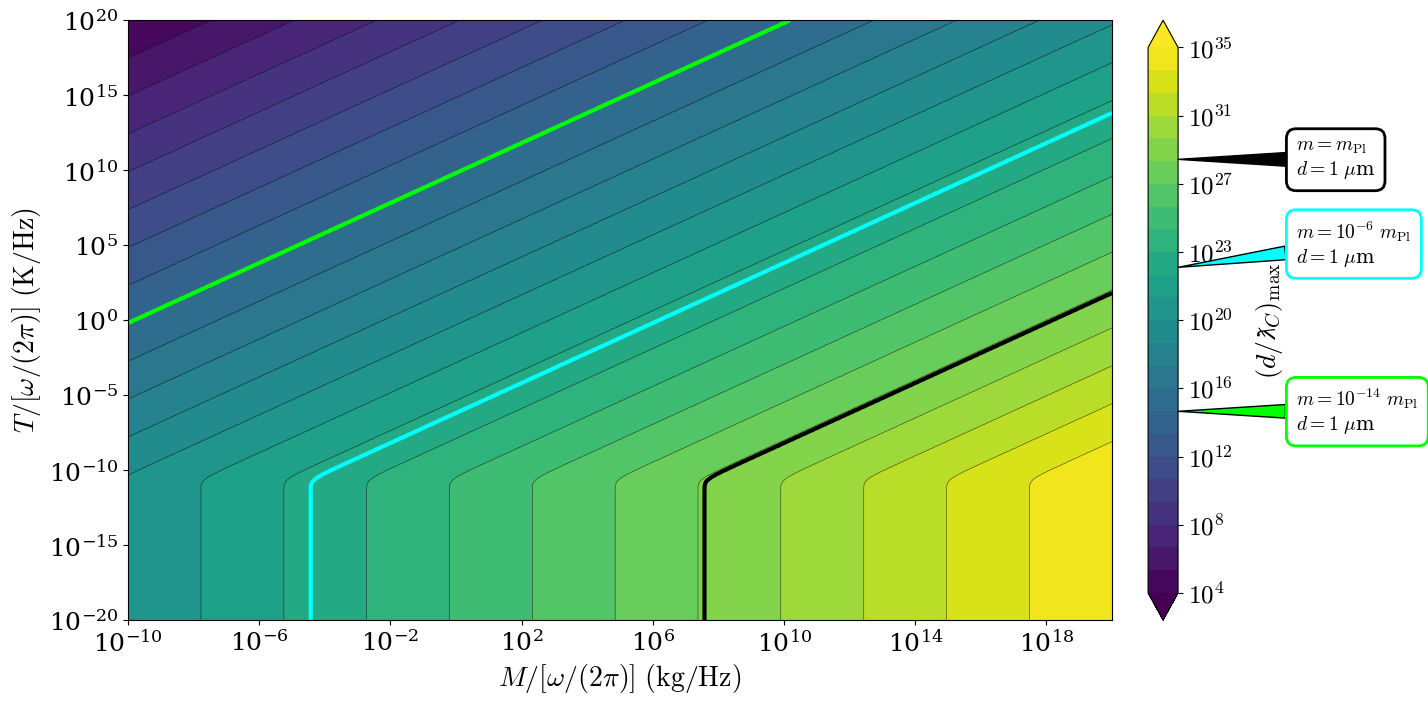}\\
 \caption{Momentum-conservation constraint on the preparation of coherent spatial superpositions. The figure shows the
 contour plot of the r.h.s.\ of (\ref{eq:main-bound-lambdaC}), which
 imposes an upper bound on the separation $d$ (in units of reduced Compton wavelength, $\lambdabar_C$) of a coherent spatial superposition of a particle of mass $m$. The figure can be used in
 two ways. Given the parameters $(M,T,\omega)$ related to the apparatus setup, they mark a point in the figure, 
which falls on top of a contour whose value can be read from the color bar. Only state separations $d/\lambdabar_C$ {\it smaller} than this value have a chance of being coherently superposed by such an apparatus. Alternatively, once
the experimentalist establishes a target value for $d/\lambdabar_C$ --- some reference cases are shown on the color  bar --- the
contour corresponding to that value is determined on the figure. Then, only apparatuses with parameters
$(M,T,\omega)$ giving a point which falls to the {\it right} of the target contour can be used to prepare such a superposition.
}
  \label{fig:bound}
\end{figure*}

In Fig.~\ref{fig:bound}, we plot the r.h.s.\ of~(\ref{eq:main-bound-lambdaC}), labeled  $(d/\lambdabar_C)_{\text{max}}$.
The figure also highlights three representative cases of particle masses for a given spatial separation
$d = 1~\mu{\text m}$. 
Current matter-wave interferometry experiments
have reached
masses of order $10^{-14}~m_\text{Pl}$~\cite{Pedalino2026}. For these masses, preparing a coherent superposition of states separated by
$d = 1~\mu{\text m}$ requires apparatuses with effective parameters $(M,T,\omega)$ related to points
in Fig.~\ref{fig:bound} lying to the right of the green contour --- a very undemanding requirement
in terms of temperature and mass
of the apparatus, rendering the constraint effectively irrelevant in this regime.
BMV-type proposals~\cite{Bose2017,Marletto2017}, with masses in the range $10^{-7}$--$10^{-6}~m_\text{Pl}$ (cyan contour), 
are substantially
 more demanding
but still 
within reach of current or near-future technology~\cite{Westphal2021}. 
By contrast, as the particle mass approaches the Planck scale  (black contour), the momentum-conservation constraint becomes increasingly restrictive and eventually excludes the preparation of coherent spatial superpositions unless the control over the (quantum) apparatus is correspondingly improved.
For the sake of illustration, for $m=m_\text{Pl}$ and 
$d=1~\mu\text{m}$, an apparatus with $M=100$~kg and $\omega/(2\pi)=1$~Hz fails the bound at {\it any} temperature, as can be
seen from both Figs.~\ref{fig:coherence_mT} and~\ref{fig:bound}. Meeting the constraint at the Planck mass would require both an extremely soft anchoring, $\omega/(2\pi) \lesssim 10^{-5}$~Hz, and cooling of the CM mode to $T\lesssim 2\times 10^{-16}$~K --- many orders of magnitude below any CM-mode temperature achieved experimentally, indicating that momentum conservation alone already imposes a demanding requirement for the realization of Planck-mass spatial superpositions. Curiously, note that Newton's gravitational constant $G_N$ appears nowhere in the analysis,
so that the Planck mass $m_\text{Pl}= \sqrt{\hbar c/G_N}$ 
plays absolutely no special role here; it is no more than an accident that superposing particles with masses near the Planck scale represents such
a challenging experimental enterprise.


It is important to emphasize that the bound in Ineq.~\eqref{eq:main-bound} [or, equivalently, ({\ref{eq:main-bound-lambdaC}})]
accounts for only a single source of decoherence: the entanglement between the particle and the CM degree of freedom of the apparatus that is unavoidably generated by momentum conservation. All other decoherence channels, including gas collisions, photon scattering, gravitational decoherence, gravitational radiation, and coupling to internal degrees of freedom of the apparatus, act independently and must be suppressed separately.

In particular, as discussed in the following, the localized impulse responsible for splitting the particle wave packet will generically excite the internal elastic degrees of freedom of the apparatus. These modes become entangled with the particle's which-path information in much the same way as the CM-DoF. Unlike the latter, however, they are typically coupled to a dense spectrum of internal degrees of freedom and therefore decohere on substantially shorter timescales through interactions with the apparatus's own phonon 
bath~\cite{Henkel2022,Folman2024}. Their effect is thus to further tighten the constraints derived above.

For this reason, the scenarios shown here should be regarded as optimistic upper bounds on what can be achieved under momentum conservation alone. Any additional source of decoherence can only make the preparation of coherent spatial superpositions more demanding than the estimates presented here.

\section{Discussion}

\vspace{0.2cm}
\noindent\textbf{What plays the role of the apparatus}

\noindent The bound in Ineq.~\eqref{eq:main-bound} [equivalently, ({\ref{eq:main-bound-lambdaC}})] depends on an apparatus mass $M$ that has been treated as a given parameter. However, in practice, no laboratory device is truly isolated. Any experimental apparatus is mechanically connected to its mount, to the building, and ultimately to the Earth itself. Therefore, we must clarify which mass should enter 
Ineqs.~\eqref{eq:main-bound} and (\ref{eq:main-bound-lambdaC}) and what the consequences of different choices are. Taken at face value, the bound suggests that increasing $M$ is always advantageous: a heavier apparatus may be hotter while still preserving coherence (e.g., moving up along the diagonal contour lines in Fig.~\ref{fig:bound}). The actual situation is more subtle, and the trade-off discussed in the following constitutes the main practical limitation associated with the present analysis.

Let $\tau_{\rm prep}$ denote the duration of the splitting interaction and $c_s$ the characteristic speed of sound in the materials that make up the apparatus. During the preparation stage, the reaction associated with the splitting propagates away from the active element of the apparatus  --- for example, a trap, a mirror, or a magnetic structure --- at approximately the speed $c_s$. Only those subsystems lying within an acoustic horizon of size $\sim c_s\tau_{\rm prep}$ and that remain mechanically rigid on the time scale $\tau_{\rm prep}$ can participate in the recoil as a single coherent body. Degrees of freedom located beyond this acoustic horizon, or connected only through sufficiently soft mechanical degrees of freedom, do not participate in the backreaction during the preparation stage. Any momentum transferred to them occurs only after the splitting process has already taken place, and they therefore act effectively as part of the environment. For typical values $\tau_{\rm prep}\sim 1~{\rm ms}$ and $c_s\sim5\times10^3~{\rm m/s}$, the corresponding rigid region extends over only a few meters, limiting the mass of the operating apparatus to that of an optical table or smaller.

Within this operational definition, it may seem natural to relax the bound in~\eqref{eq:main-bound} by enlarging the apparatus to include as much rigidly connected hardware as possible. However, this strategy may be counterproductive. The impulse required to displace the particle by $\pm\vec d/2$ is applied locally, and its redistribution through the apparatus generally excites internal degrees of freedom. These internal degrees of freedom become correlated with the particle's path information in much the same way as the CM degree of freedom. Then two distinct effects arise.

The first is purely kinematic. Because the reaction force is applied locally, rigid recoil requires the associated elastic disturbance to propagate throughout the apparatus within the preparation time $\tau_{\rm prep}$. As the apparatus grows, this condition progressively fails for more distant regions. Consequently, the effective mass participating in the coherent recoil does not increase indefinitely with the size of the assembly. Instead, it saturates the mass contained within the acoustic horizon, while the remaining degrees of freedom behave as a loosely coupled environment of internal degrees of freedom.

The second effect is more problematic. Even within the rigid region, a localized impulse generally excites a superposition of CM motion and internal elastic degrees of freedom. These excitations typically possess a large density of states, are coupled to the apparatus's internal thermal reservoir, and thermalize on timescales much shorter than those relevant to any realistic interference protocol. The which-path information encoded in such degrees of freedom is therefore effectively lost. Unlike the CM recoil considered previously, the resulting loss of coherence cannot, in general, be reversed.

The trade-off is therefore clear. Increasing $M$ relaxes the bounds in~\eqref{eq:main-bound} and (\ref{eq:main-bound-lambdaC}), but simultaneously enlarges the internal sector of the apparatus and increases the fraction of recoil that leaks into uncontrollable degrees of freedom. This leakage constitutes genuine decoherence and cannot be mitigated by lowering the temperature or modifying the trapping frequency. Conversely, decreasing $M$ tightens the bounds in~\eqref{eq:main-bound}  and (\ref{eq:main-bound-lambdaC}), but favors a situation in which the apparatus behaves approximately as a single controllable quantum degree of freedom, with its internal modes playing only a negligible role, and the CM recoil can be reversed in principle.

The optimal choice of $M$ is therefore not the largest available mass, but rather the largest mass for which the apparatus CM can still be treated, to a good approximation, as an independent quantum degree of freedom characterized by well-defined position and momentum uncertainties, while the internal modes carry only a negligible fraction of the recoil. Determining this optimum quantitatively would require a detailed model of the elastic response of the apparatus, which is beyond the scope of the present work. 
Moreover, consistency between the mass-dipole conservation argument (strictly valid for an isolated ``system'') 
and the confining potential 
(which introduces forces on the ``system'')
requires the splitting to be fast on 
the timescale of the trap, $\omega \tau_\text{prep}\ll 1$, so that the particle-apparatus pair is effectively isolated 
during the preparation; corrections are $O((\omega \tau_\text{prep})^2)$. For slower protocols, part of the recoil 
is transferred to the anchoring structure during the preparation itself, and $M$ and $\omega$ must then be 
reinterpreted along the lines discussed above about what counts as ``the apparatus.''
However, the qualitative conclusion is that the bound expressed in~\eqref{eq:main-bound}  or (\ref{eq:main-bound-lambdaC}) provides a necessary condition for coherence of the reduced particle state, but for sufficiently large $M$ it eventually ceases to be the dominant constraint. Before the temperature and confinement requirements become limiting, it is expected that decoherence associated with internal degrees of freedom takes over.

\vspace{0.2cm}
\noindent\textbf{False decoherence}

\noindent The notion of false decoherence was introduced by Unruh~\cite{Unruh2000} to describe situations in which a system appears decohered at the level of its reduced density matrix, even though the full state of the system and its environment remains pure (or only weakly mixed), so that the lost coherence is, in principle, recoverable through a suitable manipulation of the joint system. Related ideas have recently appeared in studies of horizon-induced decoherence~\cite{Danielson2025minimize}, where the apparent loss of coherence is associated with entanglement that does not irreversibly transfer information to a genuine environment.

The bound in~\eqref{eq:main-bound} naturally falls within this framework. Indeed, it is required only if the CM degree of freedom of the apparatus is traced out, that is, only if the apparatus is regarded as part of the environment. The joint state $\hat{\rho}$ of Eq.~\eqref{eq:Ent} remains a
fully correlated particle–apparatus state whose preparation is unitarily reversible. Consequently, any protocol that does not require the reduced particle state $\hat{\rho}_S$ itself to be coherent may, in principle, operate directly on the full state $\hat{\rho}$.

A particularly relevant example is provided by the BMV protocol~\cite{Bose2017,Marletto2017}. In its idealized form, two particles, $A$ and $B$, are independently prepared in spatial superposition states, allowed to interact gravitationally, and finally recombined. The signature of the protocol is the entanglement generated between the two particles during the gravitational interaction~\cite{Carney2019,Christodoulou2023}.

The analysis presented here shows that, unless the bound in~\eqref{eq:main-bound} is satisfied, the particles $A$ and $B$ are not themselves in coherent spatial superpositions during the interaction stage. Instead, each particle is entangled with the CM degree of freedom of the apparatus used to prepare it. Denoting these CM modes by $X_A$ and $X_B$, the relevant quantum systems during the gravitational phase are therefore not $A$ and $B$ alone, but rather the composite systems $A\otimes X_A$ and $B\otimes X_B$.

A natural response would be to demand that the apparatuses themselves satisfy the bound in~\eqref{eq:main-bound}, a requirement that becomes increasingly restrictive as the particle masses grow. However, an alternative is to maintain coherent control of the apparatuses throughout the entire protocol. If the same apparatus that generates the splitting is also used to perform the recombination, then the particle-apparatus entanglement created during the preparation stage can be coherently reversed. In that case, the gravitational interaction takes place between the composite quantum systems $A\otimes X_A$ and $B\otimes X_B$, and the resulting entanglement can be transferred back to the particles during recombination in exactly the same spirit as in the idealized BMV scenario.

This observation leads to an important conceptual shift~\cite{BartlettRudolphSpekkens2007}. Once the apparatus is included as part of the quantum system, conservation of the mass dipole moment implies that the relevant branch-dependent mass distribution is no longer that of the particle alone. Since the total mass dipole moment remains fixed, it is instead the superposition of  configurations of the combined 
particle-apparatus system differing only in {\it higher multipole} moments that acts as the source of gravitational entanglement. 

However, for false decoherence to provide a viable description, several conditions must be satisfied. First, the joint particle-apparatus system must remain effectively isolated from genuine environmental decoherence channels throughout the duration of the protocol. Second, the recoil associated with the splitting process must be absorbed almost entirely by the CM degree of freedom, with negligible excitation of the internal modes discussed before. Otherwise, those internal degrees of freedom become an effective environment, and the resulting loss of coherence cannot be reversed. Finally, the splitting interaction itself must be reversible in the sense that the same apparatus, prepared in the same internal state, can coherently implement both the splitting and the subsequent recombination of the particle wave packet.

If these requirements are met, then the bound in~\eqref{eq:main-bound} should be viewed, in Unruh's sense, as a false bound. Its apparent restriction arises only after tracing out the apparatus. By treating the apparatus as part of the quantum system rather than as part of the environment, the limitation imposed by momentum conservation can, in principle, be circumvented.

\vspace{0.2cm}
\noindent\textbf{Conclusions}

\noindent The main observation of this work is that the recoil induced by momentum conservation imposes a quantitative constraint on the preparation of massive spatial superpositions. Because the argument relies only on momentum conservation, the canonical commutation relations, and the finite coherence length of the apparatus center-of-mass state, the resulting bound is universal, being independent of the microscopic details of the preparation protocol. In this sense, it constitutes a model-independent limitation that applies to any scheme in which a massive particle is placed in a coherent spatial superposition by means of a finite-mass apparatus. Although the effect is entirely negligible for present-day matter-wave interferometry experiments, it becomes progressively more restrictive as one approaches the mass scales currently considered in tests of gravitationally mediated entanglement and in proposals aimed at probing quantum mechanics near the Planck scale. For the representative parameters considered here, momentum conservation alone already demands apparatus temperatures far below those currently achievable when approaching Planck-mass superpositions  --- although, as discussed at the end of 
Sec.~I, the Planck mass itself plays no fundamentally distinct 
role in the discussion. This observation is particularly relevant in the context of proposed tests of an objective Heisenberg cut on the Planck scale~\cite{AguiarMatsas2024,AguiarMatsas2026}. Indeed, our results show that a high suppression of coherence may arise from a completely conventional quantum-mechanical mechanism, namely the entanglement between the particle and the finite-mass apparatus required to prepare the superposition. Consequently, any claim of evidence for an objective limitation of the superposition principle at large masses must first disentangle such fundamental effects from the model-independent constraints imposed by momentum conservation itself.

An important aspect of our analysis is that it isolates what is arguably the most optimistic scenario compatible with momentum conservation. Throughout the article, we have assumed that the recoil associated with the preparation process is completely absorbed by the center-of-mass degree of freedom of the apparatus. Realistic devices inevitably possess internal vibrational modes that can also become entangled with the particle and generally decohere on much shorter timescales through coupling to their own thermal environment. The constraints derived here should therefore be interpreted as necessary, but not sufficient, conditions for the successful preparation of coherent spatial superpositions. In particular, the center-of-mass bound derived in this work provides a lower bound on the experimental cost of preparing massive superpositions: any realistic treatment of internal degrees of freedom can only strengthen the restrictions obtained here. For this reason, we expect the present results to serve both as a benchmark against which more detailed apparatus-specific analyses may be compared and as a reference point for future experimental efforts aimed at probing the limits of quantum superposition at increasingly large mass scales.

Several directions naturally emerge from the present analysis. The most immediate is the incorporation of the momentum-conservation-induced particle--apparatus entanglement into concrete interferometric protocols, particularly those aimed at probing gravitationally mediated entanglement. In such scenarios, the preparation stage already produces correlations between the particles and the devices used to manipulate them, raising the question of how these correlations modify the entanglement ultimately generated during the protocol. A second important extension is the development of microscopic models of realistic apparatuses, including their vibrational degrees of freedom. Such an analysis would allow one to determine when the center-of-mass constraint derived here constitutes the dominant limitation and when it is superseded by decoherence associated with internal modes. Finally, because the underlying mechanism relies only on conservation laws and quantum correlations, it is natural to ask whether analogous constraints arise in relativistic settings, where the relevant conserved quantity is the stress-energy tensor rather than momentum alone. We hope that the present work stimulates further investigation of these questions and contributes to a more complete understanding of the physical requirements for preparing increasingly massive quantum superpositions.

\section*{Acknowledgments}
 
D.V.\ thanks Gerard Higgins for valuable discussions about experimental aspects of  spatial superposition of massive particles  during a sabbatical leave at the Institute for Quantum Optics and Quantum Information (IQOQI-Vienna), supported by the S\~ao Paulo State Research Foundation (FAPESP), under grant no.\ 2023/04827-9. L.C.C.\ acknowledges support from the National Council for Scientific and Technological Development (CNPq) through grant 308065/2022-0, the National Institute of Science and Technology for Applied Quantum Computing through CNPq grant 408884/2024-0, and Goiás State Research Foundation (FAPEG) through grant 202510267001843. D.O.S.P.\ acknowledges support from the National Council for Scientific and Technological Development (CNPq) through grant 304891/2022-3. D.O.S.P and L.C.C. acknowledge support from FAPESP through grant 2025/23726-4.


\begin{thebibliography}{99}


\bibitem{Bassi2013} A.\ Bassi, K.\ Lochan, S.\ Satin, T.\ P.\ Singh, H.\ Ulbricht, {\it Models of wave-function collapse, underlying theories, and experimental tests}, Rev.\ Mod.\ Phys.\ \textbf{85}, 471 (2013).

\bibitem{Bassi2023} A.\ Bassi, M.\ Dorato, and H.\ Ulbricht, {\it Collapse models: A theoretical, experimental and philosophical review}, Entropy \textbf{25}, 645 (2023).

\bibitem{Bose2017}
S.~Bose~\textit{et al.},
\emph{Spin entanglement witness for quantum gravity},
Phys.\ Rev.\ Lett.\ \textbf{119}, 240401 (2017).

\bibitem{Marletto2017}
C.~Marletto and V.~Vedral,
\emph{Gravitationally-induced entanglement between two massive particles is
sufficient evidence of quantum effects in gravity},
Phys.\ Rev.\ Lett.\ \textbf{119}, 240402 (2017).



\bibitem{Carney2019}
D.~Carney, P.~C.~E.~Stamp, and J.~M.~Taylor,
\emph{Tabletop experiments for quantum gravity: a user's manual},
Class.\ Quantum Grav.\ \textbf{36}, 034001 (2019).

\bibitem{Degen2017} C.\ L.\ Degen, F.\ Reinhard, and P.\ Cappellaro, {\it Quantum sensing}, Rev.\ Mod.\ Phys.\ \textbf{89}, 035002 (2017).

\bibitem{Isart2010} O.\ Romero-Isart, M.\ L.\ Juan, R.\ Quidant, and J.\ I.\ Cirac, {\it Toward quantum superposition of living organisms}, New J.\ Phys.\ \textbf{12}, 033015 (2010).



\bibitem{RomeroIsart2011}
O.~Romero-Isart \textit{et al.},
\emph{Large quantum superpositions and interference of massive
nanometer-sized objects},
Phys.\ Rev.\ Lett.\ \textbf{107}, 020405 (2011).


\bibitem{Henkel2022}
C.~Henkel and R.~Folman,
\emph{Internal decoherence in nano-object interferometry due to phonons},
AVS Quantum Sci.\ \textbf{4}, 025602 (2022).

\bibitem{Folman2024}
Y.~Japha, T.~Grover, C.~Henkel, and R.~Folman,
\emph{Universal limit on spatial quantum superpositions with massive
objects due to phonons},
Phys.\ Rev.\ A \textbf{110}, 042221 (2024).

\bibitem{Satishchandran2024}
G.~Satishchandran and R.~M.~Wald,
\emph{The asymptotic behavior of massless fields and the memory effect},
Phys.\ Rev.\ D \textbf{99}, 084007 (2019).

\bibitem{Gunnink2023}
M.~Gunnink, A.~Mazumdar, M.~Schut, and M.~Toro\v{s},
\emph{Gravitational decoherence by the apparatus in the
quantum-gravity-induced entanglement of masses},
Class.\ Quantum Grav.\ \textbf{40}, 235006 (2023).

\bibitem{Grossardt2020}
A.~Gro{\ss}ardt,
\emph{Acceleration noise constraints on gravity-induced entanglement},
Phys.\ Rev.\ A \textbf{102}, 040202(R) (2020).

\bibitem{BohrEinstein}
N.~Bohr,
\emph{Discussion with Einstein on epistemological problems in atomic
physics}, in
\emph{Albert Einstein: Philosopher--Scientist}, edited by P.~A.~Schilpp,
The Library of Living Philosophers, Vol.~7
(Open Court, La Salle, IL, 1949), pp.\ 200--241.

\bibitem{WoottersZurek1979}
W.~K.~Wootters and W.~H.~Zurek,
\emph{Complementarity in the double-slit experiment: Quantum
nonseparability and a quantitative statement of Bohr's principle},
Phys.\ Rev.\ D \textbf{19}, 473 (1979).

\bibitem{Liu2024recoiling}
X.-J.~Liu \emph{et al.},
\emph{Tunable Einstein--Bohr recoiling-slit gedankenexperiment at the
quantum limit},
Phys.\ Rev.\ Lett.\ {\bf 135}, 230202 (2025).

\bibitem{JoosZeh1985}
E.~Joos and H.~D.~Zeh,
\emph{The emergence of classical properties through interaction with the
environment},
Z.\ Phys.\ B \textbf{59}, 223 (1985).


\bibitem{Baumgratz2014} T.\ Baumgratz, M.\ Cramer, and M.\ Plenio, {\it Quantifying coherence}, Phys. Rev. Lett. \textbf{113}, 140401 (2014).


\bibitem{Feynman}
R.\ P.\  Feynman,  {\it Statistical Mechanics: A Set of Lectures} (W. A. Benjamin,  Reading,  MA, 
1972).


\bibitem{Pedalino2026} S.\ Pedalino, B.\ E.\ Ramírez-Galindo, R.\ Ferstl, K.\ Hornberger, M.\ Arndt, and S.\ Gerlich, {\it Probing quantum mechanics with nanoparticle matter-wave interferometry}, Nature \textbf{649}, 866 (2026). 



\bibitem{Westphal2021}
T.~Westphal, H.~Hepach, J.~Pfaff, and M.~Aspelmeyer,
\emph{Measurement of gravitational coupling between millimetre-sized
masses},
Nature \textbf{591}, 225 (2021).

\bibitem{Unruh2000}
W.~G.~Unruh,
\emph{False loss of coherence}, in
\emph{Relativistic Quantum Measurement and Decoherence},
edited by H.-P.~Breuer and F.~Petruccione,
Lect.\ Notes Phys.\ \textbf{559}, 125 (Springer, Berlin, 2000);
arXiv:quant-ph/0008045.



\bibitem{Danielson2025minimize}
D.~L.~Danielson, G.~Satishchandran, and R.~M.~Wald,
\emph{Killing horizons decohere quantum superpositions},
Phys.\ Rev.\ D \textbf{108}, 025007 (2023);
\emph{Local description of decoherence of quantum superpositions by black
holes and other bodies}, Phys.\ Rev.\ D {\bf 111}, 025014 (2025).



\bibitem{Christodoulou2023}
M.~Christodoulou \textit{et al.},
\emph{Locally mediated entanglement in linearized quantum gravity},
Phys.\ Rev.\ Lett.\ \textbf{130}, 100202 (2023).



\bibitem{BartlettRudolphSpekkens2007}
S.~D.~Bartlett, T.~Rudolph, and R.~W.~Spekkens,
\emph{Reference frames, superselection rules, and quantum information},
Rev.\ Mod.\ Phys.\ \textbf{79}, 555 (2007).

\bibitem{AguiarMatsas2024}
G.~H.~S.~Aguiar and G.~E.~A.~Matsas,
\emph{A simple gravitational self-decoherence model},
Phys.\ Rev.\ D {\bf 112}, 046004 (2025).
 
\bibitem{AguiarMatsas2026}
G.~H.~S.~Aguiar and G.~E.~A.~Matsas,
\emph{Will we ever quantize the center of mass of macroscopic systems? A case for a Heisenberg cut in quantum mechanics},
Braz.\ J.\  Phys.\ {\bf 56}, 192 (2026).





\end{thebibliography}
\end{document}